\documentclass{sig-alternate}

\begin{document}
%

\title{On the random access performance of Cell Broadband Engine with graph analysis application
}

\numberofauthors{3} 
\author{
\alignauthor
Mingyu Chen\\
       \affaddr{Key Laboratory of Computer System and Architecture}\\
       \affaddr{Inst. of Comp. Tech., Chinese Academy of Sciences}\\
       \affaddr{P.O.Box 2704}\\
      \affaddr{Beijing,100190,China}\\
       \email{cmy@ict.ac.cn}
\alignauthor
David A. Bader\\
      \affaddr{College of Computing}\\
       \affaddr{Georgia Institute of Technology}\\
       \affaddr{Atlanta, GA 30332,United States}\\
       \email{bader@cc.gatech.edu}
}

\maketitle
\begin{abstract}
The Cell Broad Engine (BE) Processor has unique memory access architecture besides its powerful computing engines. Many computing-intensive applications have been ported to Cell/BE successfully. But memory-intensive applications are rarely investigated except for several micro benchmarks. Since Cell/BE has powerful software visible DMA engine, this paper studies on whether Cell/BE is suit for applications with large amount of random memory accesses. Two benchmarks, GUPS and SSCA\#2, are used. The latter is a rather complex one that in representative of real world graph analysis applications. We find both benchmarks have good performance on Cell/BE based IBM QS20/22. Compared with 2 conventional multi-processor systems with the same core/thread number, GUPS is about 40-80\% fast and SSCA\#2 about 17-30\% fast. The dynamic load balancing and software pipeline for optimizing SSCA\#2 are introduced.  Based on the experiment, the potential of Cell/BE for random access is analyzed in detail as well as its limitations of memory controller, atomic engine and TLB management.Our research shows although more programming effort are needed, Cell/BE has the potencial for irregular memory access applications.
\end{abstract}

\category{D.1.3}{Programming Techniques}
{Concurrent Programming}{- Parallel programming}
 \category{C.4}{Performance of Systems}{Design studies}

\terms{programming,performance}

\keywords{Cell/BE, Random Access}

\section{Introduction}
Multi-core and many-core architectures have been widely investigated in recent days. The Cell Broadband Engine (Cell/BE) \cite{IBM_CELLBE} is a unique architectural multi-core design by Sony, Toshiba, and IBM (STI). There have been a lot of studies on computing-intensive applications on Cell/BE. Though primarily targeting high performance multimedia and gaming application, the Cell/BE has a unique memory architecture compared with convention multi-core CPU. Cell/BE has a 204GB/s internal bus and 25.6GB/s main memory access bandwidth. More specially Cell/BE allows the program to fully control the memory access via explicitly DMA operations. Total 128 DMA operations may exist simultaneously in theory.

At the same time, there are large collections of applications with randomly memory access behaviors such as graph exploration \cite{HPC_Graph_analysis,David_peta}. This kind of applications is not suitable for the conventional cache-based multi-core processors. In such applications, the data set is much larger than the processor cache and the data access pattern are nearly random with neither temporal locality nor spatial locality. The computation ratio is normally small compared with the memory access overhead, which leaves the most powerful FPUs in modern processors useless.	

A common myth about the Cell/B.E.'s memory subsystem is that it is inadequate for irregular data accesses due to the software intervention
in the memory access mechanism. Yet, this additional increase (few instructions) is relatively small compared to the hundred cycles or even
more DRAM access latency. Also, as the Cell/BE enables fine-grained control over data transfer, we can apply multiple techniques to hide the memory access latency.

In this paper, we investigate if the unique design of memory system in Cell/BE was suit for memory-intensive applications. Previous works have studied on certain kernel applications. \cite{kistler_cell_2006} gave a completely micro benchmark on communication network of Cell/BE. \cite{bader_design_2007} implemented list-ranking using software managed thread. \cite{villa_challenges_2007} presented a lock-free BFS algorithm utilizing the Cell/BE on-chip memory for bitmap. \cite{chow_Largefft} studied on large FFT over Cell/BE. However, all these applications are rather simple kernels than real world applications.	

Our study is based on two public benchmarks also. One is GUPS \cite{GUPS_web}, which is part of the HPC Challenge benchmark suite; the other is SSCA2 benchmark \cite{SSCA_2_Spec,HPC_Graph_analysis}, which is one of the HPCS Scalable Synthetic Compact Applications previously. The GUPS is a pure exhaustive random access benchmark kernel. Its performance is given by Giga Updates Per Seconds. We use it to evaluate the capability of the Cell/BE memory system. The SSCA\#2 is a relative complex benchmark, which came from real word graph analysis applications include network analysis, data mining and computational biology etc. SSCA\#2 computes the betweenness centrality of each vertex in a weighted directed graph. The performance metric is Traversed Edges Per Second (TEPS). The algorithm we used was proposed in \cite{ulrik_brandes_faster_2001,bader_design_2005}, which is in fact a BFS flow associated with stateful and coherent data structure.
 	
We have implemented both benchmarks for Cell/BE with detailed experimental evaluation on IBM QS 20(and QS22) Cell/BE blade. Overall results show that Cell/BE is 17\% -80\% faster than  traditional cache-based multi-core SMP system with the same core/threads and near memory bandwidth.  Our work demonstrates that Cell/BE has the potential to deal with complex memory-intensive applications.

Our main contributions are summarized here:
\begin{itemize}
\item We get a 0.062 GUPS on QS 20, which is more than 40-80\% higher compared with 2 16 core/thread conventional  multi-core system.
\item We show that the Cell/BE DMA-list mechanism has even more potential for random access. Only 2 of the 16 SPEs will reach 97\% of the peak performance.
\item We find the Cell/BE TLB update mechanism affect the performance greatly. The performance nearly doubled after adopting huge-TLB configuration. 
\item Using dynamic load balancing and software pipeline mechanism, we achieve a 65.8M TEPS for the SSCA\#2 benchmark, which is about 17-30\% faster than  conventional multi-core system. 
\item By profiling the SSCA\#2 implementation, we find the atomic operations occupied the most time delay that limited Cell/BE to get even better result.
\end{itemize}

The remainder of this paper is organized as follows. Section 2 gives a brief overview of the Cell/BE and QS Blade memory system as well as the GUPS and SSCA\#2 benchmarks. Section 3 describes our GUPS implementation with detailed experiments to evaluate the maximum random access performance of Cell/BE. Section 4 presents our techniques in implementing SSCA2 on Cell/BE. Section 5 are test and profiling results of the SSCA\#2 implementation. Section 6 we compares the related works. Section 7 concludes the paper.

\section{The Cell/BE architecture, GUPS and SSCA2 benchmark}
\subsection{The architecture of IBM BladeCenter QS20/22}
Cell Broadband Engine is well known as a heterogeneous multi-core chip \cite{IBM_CELLBE}. It consists one traditional general-purpose 64-bit PowerPC core (PPE) and eight 128-bit SIMD coprocessor cores (Synergistic Processor Element, SPE).  All nine cores are connected via a high bandwidth bus called Element Interconnect Bus (EIB) and share coherent main memory. The IBM BladeCenter QS20 and QS22 Blades are dual-processor system implementation based on Cell/BE.

\begin{figure}
\centering
\epsfig{file=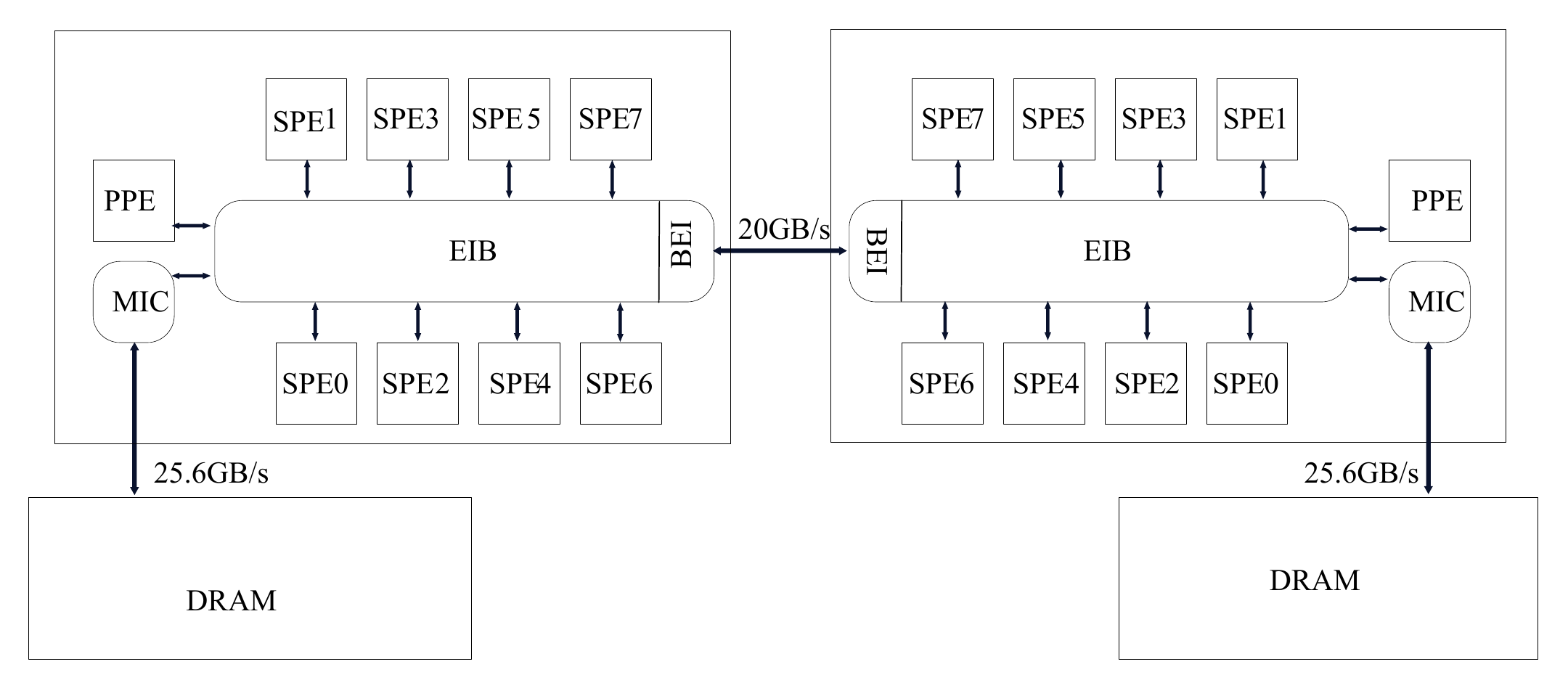,width=3.5in}
\caption{QS20/22 memory architecture}
\end{figure}
Figure 1 gives an outline of QS20/22 memory architecture.
Each Processor has a memory controller with bandwidth of 25.6GB/s. The two processors are interconnected using FlexIO interface running the fully coherent Broadband Interface (BIF) protocol. The bandwidth between two processors is 20GB/s.  As seen from the programmer, the QS blade simply consists 16 shared-memory SPEs and 2 PPEs. 

The main difference between the QS20 and QS22 is the external memory. The QS20 is configured with 1GB of XDR (Rambus) memory, while the QS22 using DDR2 SDRAM up to 32 GB. In section 4 we will show that the XDR memory has a little better random access performance than DDR2 version. 

Each SPE consists of a synergistic processor unit (SPU) and a memory flow controller (MFC). The SPE has no local cache but a 256 KB high performance local storage.  SPU core accesses data only from local storage. All external memory access and communications with other cores are through the MFC. The MFC includes a DMA controller, a memory management unit (MMU), and an atomic unit for synchronization. 

The MFC DMA controller can queue up to 16 DMA operations at the same time. The operation can be either a single DMA or a scattered DMA-list. So the whole system can support more than 250 outstanding memory operations. Each MFC also has an atomic unit that handles atomic operation, but only one reservation at a time is allowed. By default virtual memory is managed by hardware, each MFC has a 256-entry TLB with default 4KB page size.

We will see in Section 3 that the DMA queue brings more power than the memory controllers can support, while the limited TLB page size affect performance greatly.

\subsection{The Random Access benchmark (GUPS) }

The Random Access test is part of the HPC Challenge benchmark \cite{HPCC_challenge} developed for the HPCS program. The test intended to exercise the GUPS capability of a system. 

GUPS is a measurement that profiles the memory architecture of a system and is a measure of performance similar to MFLOPS. GUPS is calculated by identifying the number of memory locations that can be randomly updated in one second, divided by 1 billion. 

The basic Random Access benchmark definition \cite{GUPS_web} is:
Let $T[]$ be a table of size $2^{n}$. Let ${A_i}$ be a stream of 64-bit integers of length $2^{n+2}$ generated by the primitive polynomial over GF(2) , $X^{63} + X^{2} + X+1$.
For each $a_{i}$, set $T[a_{i} \langle 63, 64-n \rangle ] = T[a_{i} \langle 63, 64-n \rangle] + a_{i}$
Where  '$+$' denotes addition in GF(2). $a_{i} \langle l,k \rangle$denotes the sequence of bits within $a_{i}$.

The parameter $n$ defined such that: $n$ is the largest power of 2 that is less than or equal to half of main memory. The look ahead and storage before processing on distributed memory multi-processor systems is limited to 1024 per process. A small percentage of error (not exceed 1\%) is allowed for parallelization.

GUPS is good candidate for evaluating the random memory performance of a system. The process is too compact to allow further program optimization. We use GUPS as a micro benchmark tool for our study first.

\subsection {The HPCS Scalable Synthetic Compact Applications graph analysis \#2} 	 

The SSCA benchmark suite is part of DARPA High Productivity Computing Systems (HPCS) program. These benchmarks aimed to be complements to current scalable micro-benchmarks and complex real applications. SSCA\#2 is a graph theoretic problem, which is representative of computations in the field of social network, computational biology and data mining etc. 

Our study is based on SSCA\#2 v2.2 \cite{SSCA_2_Spec} specification and the C/OpenMP implementation \cite{ulrik_brandes_faster_2001}. SSCA2 contains one scalable graph generator and four computing kernels. The scalable graph generator generates a power-law scale-free graph for the computing kernels. The computing kernels all require irregular access to the graph's data structure. Since Kernel 1-3 are relatively simple and the similar computation are already included in kernel 4, we focus on Kernel 4 in our research.

Kernel 4 computes the betweenness centrality of all vertexes in a weighted directed graph. Consider a graph $G = (V,E)$, where $V$ and $E$ is the set of vertices and edges respectively.

Let $\sigma_{st}$ denote the number of shortest paths between vertices $s$  and $t$, and $\sigma_{st}(v)$ the number of those paths passing through $v$.  Betweenness Centrality of a vertex $v$ is defined as 
\begin{equation}
BC(v)=\sum_{s \ne t \ne v \in V}\frac {\sigma_{st}(v)}{\sigma_{st}}
\end{equation}

In the SSCA2 2.2.1 reference implementation, the algorithm is following the method of Brandes \cite{ulrik_brandes_faster_2001}. Brandes algorithm computes $\delta_s(v)$ using a Breadth-first search (BFS) process for each vertex $s$
\begin{equation}
\delta_s(v)=\sum_{w:v \in pred(s,v)} \frac {\sigma_{sv} } {\sigma_{sw}}\left( 1 + \delta_s \left( w \right) \right)
\end{equation}
                                
Where $pred(s,v)$ denote the predecessor set of vertex $v$ on shortest paths from $w$. Then $BC(v)$ can be obtained by sum up all $\delta_s(v)$.
                                
To compute $\delta_s$, a BFS and a back trace process are needed. In the BFS search process, besides the access sequence of each vertex, the predecessor set are also recorded, the depth $d_s(v)$ and $\sigma_{st}(v)$ are computed throughout the process.  For each vertex, the computation of $\sigma_{st}(v)$ is a multi-source adding operation and the computation of predecessor set $pred(s,v)$ is a multi-source joining operation. These two global operations bring more difficulties for parallelization than the original BFS algorithm. We will see in section 5 the atomic operations are the main obstacle for higher efficiency.
                                
The back trace process just uses the result generated during BFS and compute recursively. This process can be done in parallel without contention. But it still needs to visit all the browsed edges, which means large amount of random memory accesses. 

\section{Analyze the Cell/BE memory engine with GUPS}

Since GUPS is a simple but exhaustive random access kernel, we use it as a tool to evaluate the DMA performance of Cell/BE.

The parallelization of GUPS is straightforward:  just split the $T[]$ array equally to different threads. Since Cell/BE does not support threads within SPU, we use a multi-queue method to implement GUPS. In each SPU, we maintain multiple independent queue. For each queue, we assign a fixed-length DMA-list and keep looping  get a trunk of random numbers by a DMA-list operation, do updating, write it back to main memory, then get next trunk in sequence. 

The SPU query each thread in turn, once a DMA-list operation finished, it will be processed immediately until the following DMA operation is started and SPU came back to the query loop again. 

Three parameters are considered during the test:  queue numbers within a single SPU, DMA-list queue length for each queue, the number of SPUs.  

All results are obtained from QS20 with IBM Cell/BE SDK 3.1, Linux 2.6.25 under 16MB (huge) TLB page size unless otherwise stated. The QS20 has 1GB memory, so we did all experiments over a 512MB data size for comparison. It should be noticed that larger data size would decrease the GUPS a little.

\subsection{Single SPU test}
First, we try to figure out the best random access performance of single SPU. 

We vary queue number and queue length. As in Figure 2, for single SPU we can get the maximum of GUPS 0.0294. In fact it is about 47\% of the maximum we can ever get from multi-SPUs. The performance improves as queue number increase. However there are only a little difference when queue number large than 4, normally queue number 8 will reach maximum. The larger queue length also brings better performance but with a asymptotical improvement. 

\begin{figure}[h]
\centering
\epsfig{file=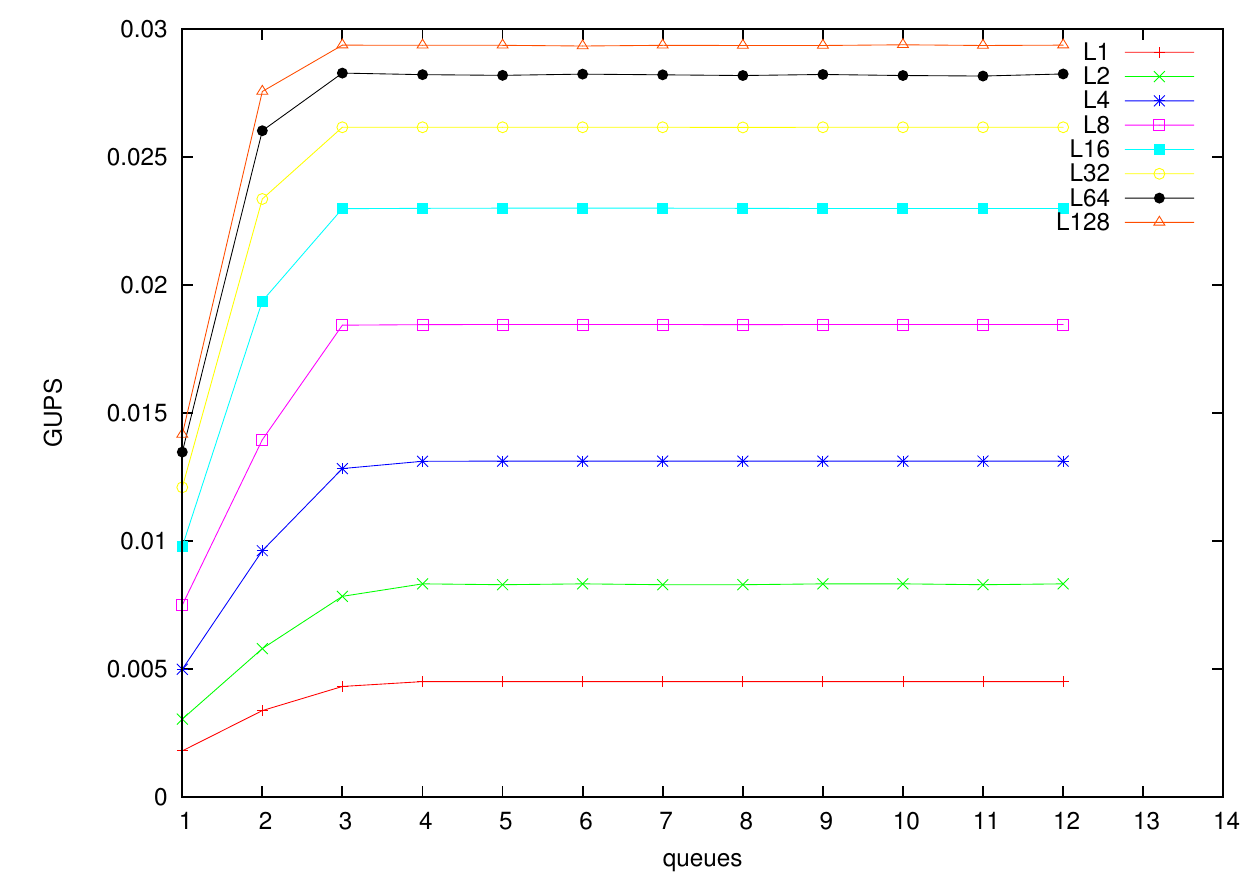,width=3.5in}
\caption{Single SPU test, varying  queue numbers and queue length}
\end{figure}

Figure 3 and 4 give the GUPS results of 2 and 4 SPUs. We can see the performance with 2 SPUs is nearly doubled. In fact it can reach nearly 97\% the maximum already. The 4-SPU result shows the peak was reached easily even with a short queue length 8-16.

\begin{figure}[h!]
\centering
\epsfig{file=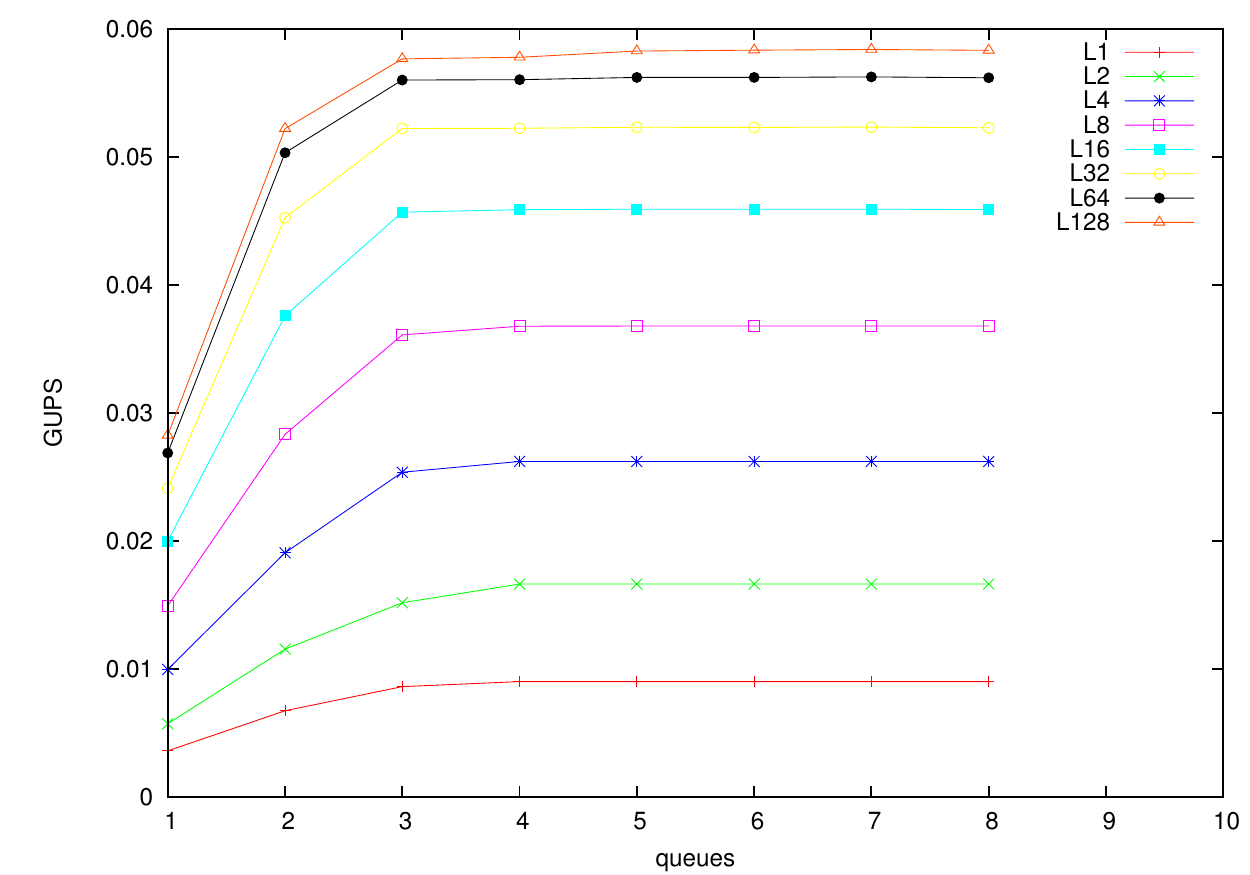,width=3.5in}
\caption{2-SPU test, varying  queue number and queue length}
\epsfig{file=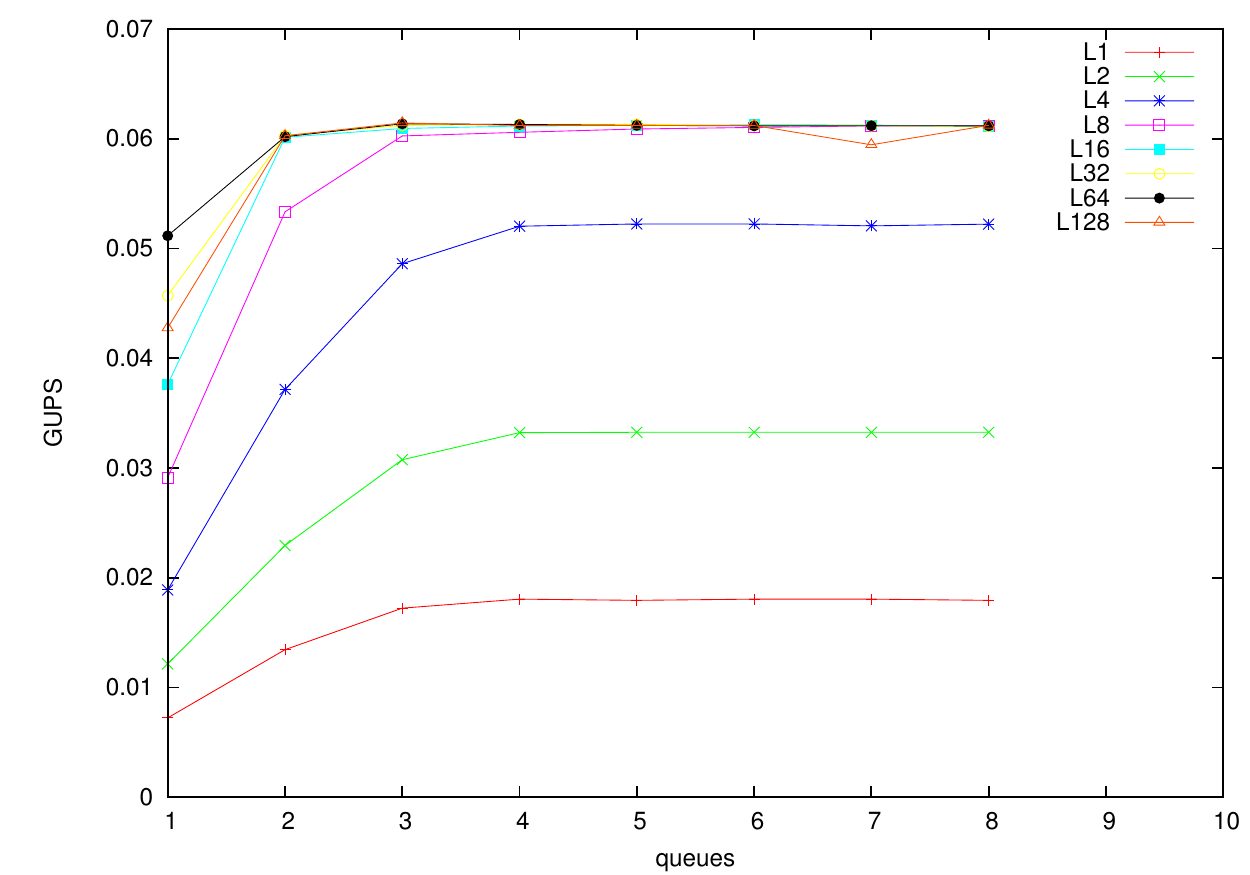,width=3.5in}
\caption{4-SPU test, varying  queue number and queue length}
\epsfig{file=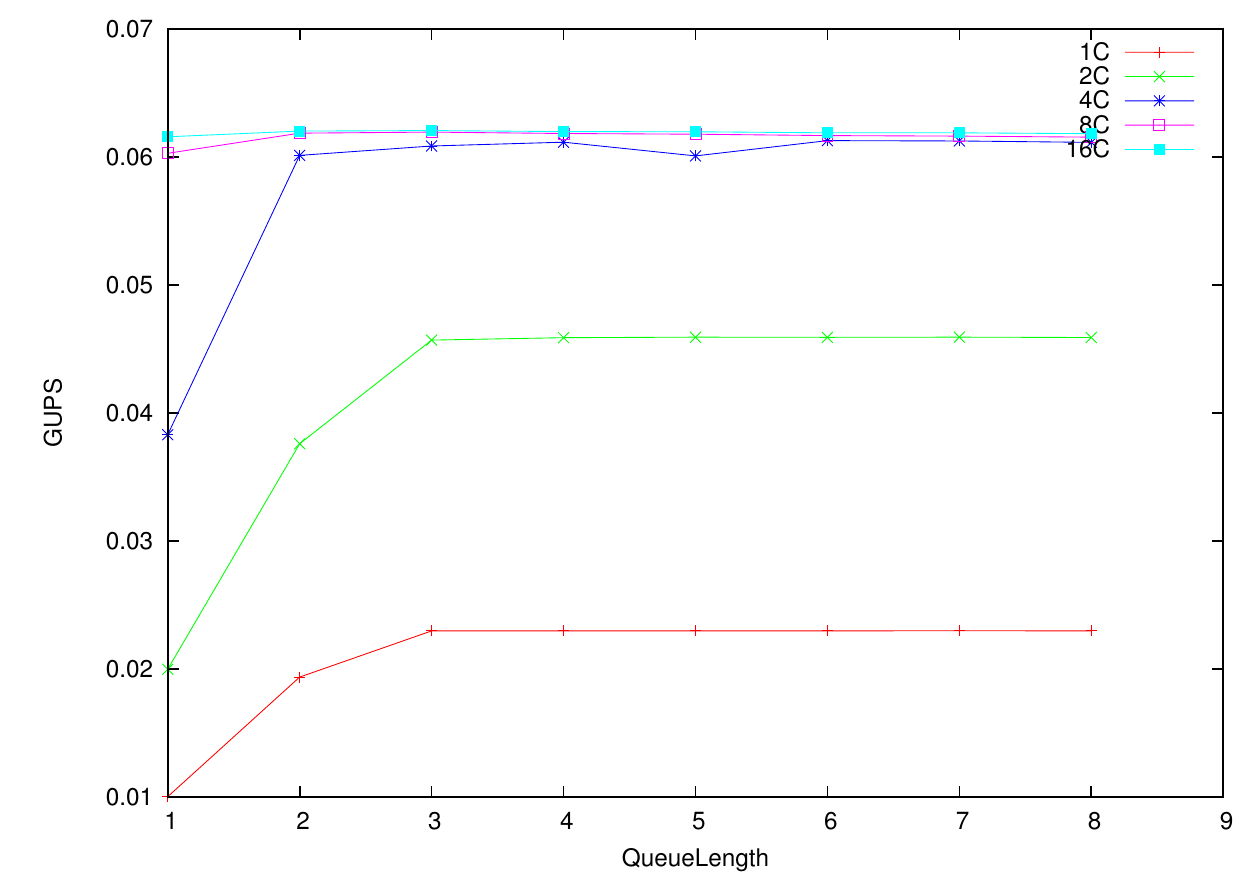,width=3.5in}
\caption{fix  queue number = 4, varying queue length and SPU number }
\end{figure}

Next we fixed queue number at 4 then varying queue length and SPU number, as Figure 5.We can see increasing SPU numbers does not increase the GUPS after 4, but needs shorter queue length. 16 SPUs can reach the peak even with queue length 1. The maximum GUPS is 0.062, which can be reached in many configurations.

With the above results, we can draw a conclusion that the Cell/BE SPU has a great potential for random access. The memory controller is the bottleneck for more GUPS. We can infer that if Cell BE were equipped more memory channels the GUPS would easily increased.

\subsection{The effect of TLB page size}
By default the Cell/BE use hardware managed TLB. The page size is 4KB. Each SPU have a 256 entries TLB table. So once the memory data size is larger than 1MB, random access will cause TLB miss and reload frequently, which has relatively larger overhead for Cell/BE. This confused us much at the early stage of the work. The effect can be viewed from figure 6,7

\begin{figure}[h]
\centering
\epsfig{file=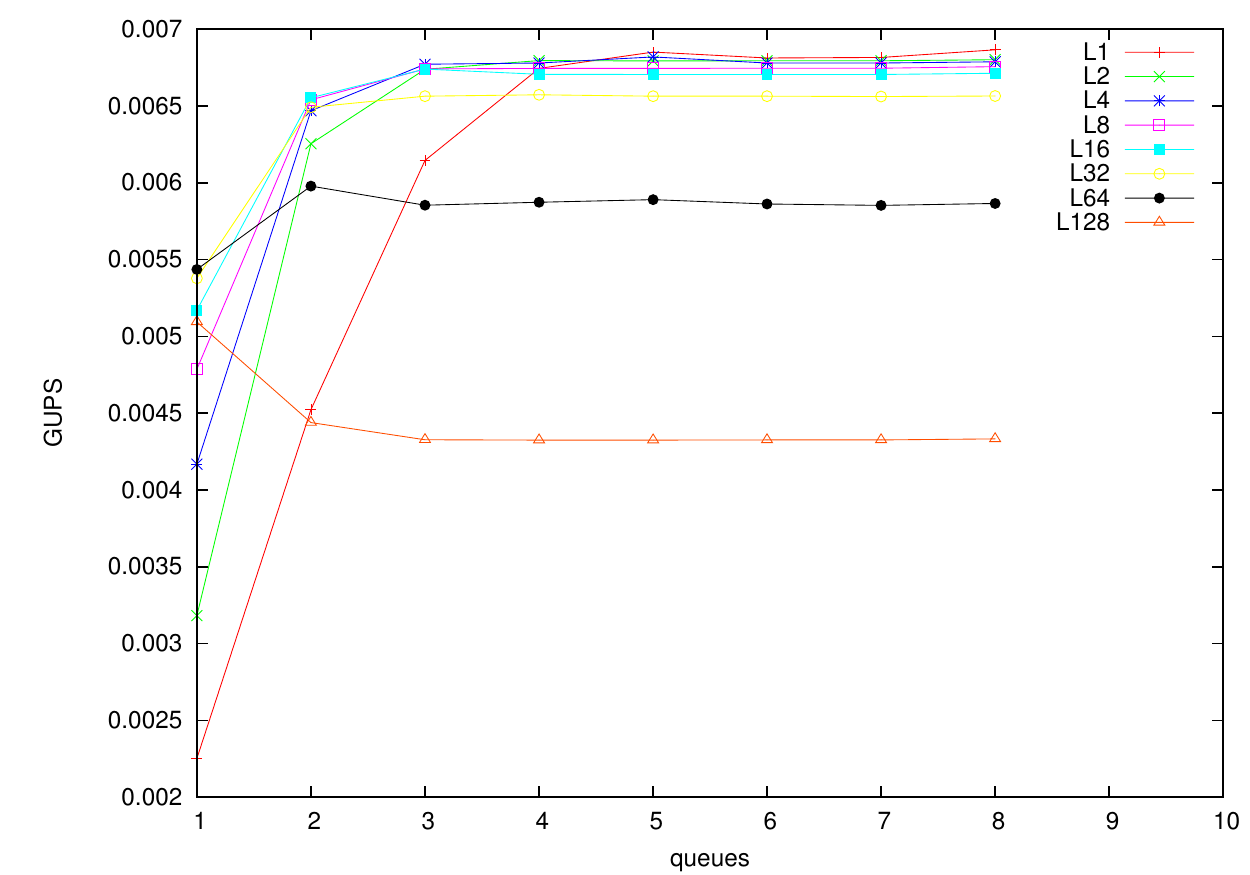,width=3.5in}
\caption{GUPS with 4K page, 1-SPU,varying queue number and queue number}
\epsfig{file=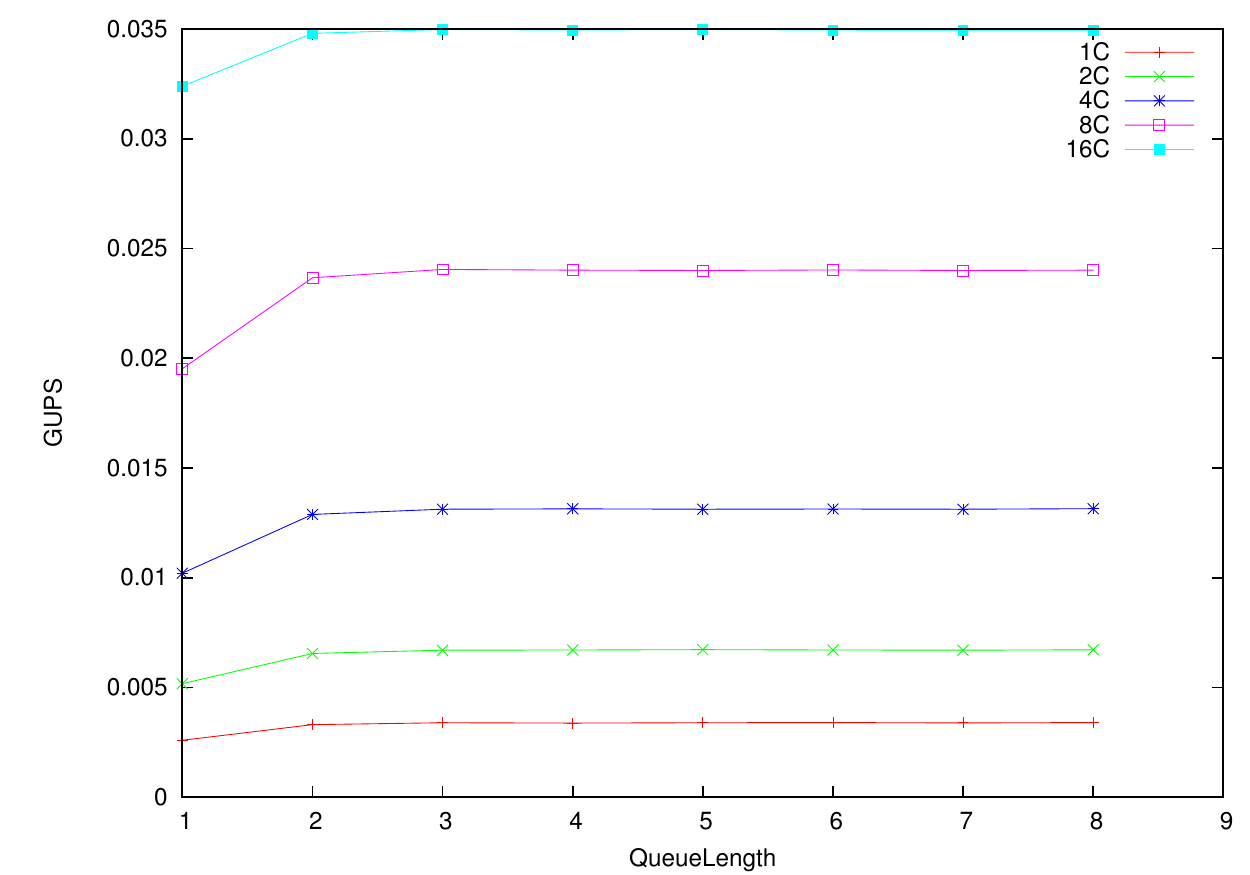,width=3.5in}
\caption{GUPS with 4K page, queue num=4,varying queue length and SPU number }
\end{figure}

From Figure 6, we can see for one SPU the peak performance is only about 20\% of previous result. A strange phenomenon is that larger queue length will get even worse result. In figure 7, 4 SPUs do not saturate the bus any more. With all 16 SPUs the performance can only reach about 56\% of the peak of HugeTLB case. We can draw a conclusion that TLB page size has a large influence on the application with random memory accesses.  

\subsection{Comparison over different platforms}
We compared 4 platforms. One is IBM QS20 which has 1GB XDR Ram, another is the newer IBM QS22 which has 32G DDR2 SDRAM, two dual 128-bit DDR2-800M memory channels.  
The third platform 'Opteron' is a quad processor SMP using AMD 4-core Opteron 8347.  Each core has 512KB L2 cache, 1K TLB entries, running at 1.9Ghz. Each processor has a shared 2MB L3 cache.   It has 4 dual-channel DDR2 memory controllers, the same as QS22 but a lower 533MHz. Bandwidth between processors is 8GB/s. The last platform 'Nehalem' is a dual processor SMP using the latest Intel  4-core Xeon 5530. Each core has 256KB L2 cache, two hyper-thread, running at 2.4GHz. Each processor has 8MB shared L3 cache. It has 2 dual-channel 1333MHz DDR3 memory. This platform has total 16 physical threads. 

We use the C/OpenMP reference implementation for X86\_64 platforms, compilers are PGI 7.2 and ICC 11.0. To compare, we also used huge TLB (2MB).  

For all platforms, we use "numactl" \cite{numactl} utility to make sure the data spread on all memory channels. We use queue number 4 and queue length 16 for all test.
 
\begin{figure}[h]
\centering
\epsfig{file=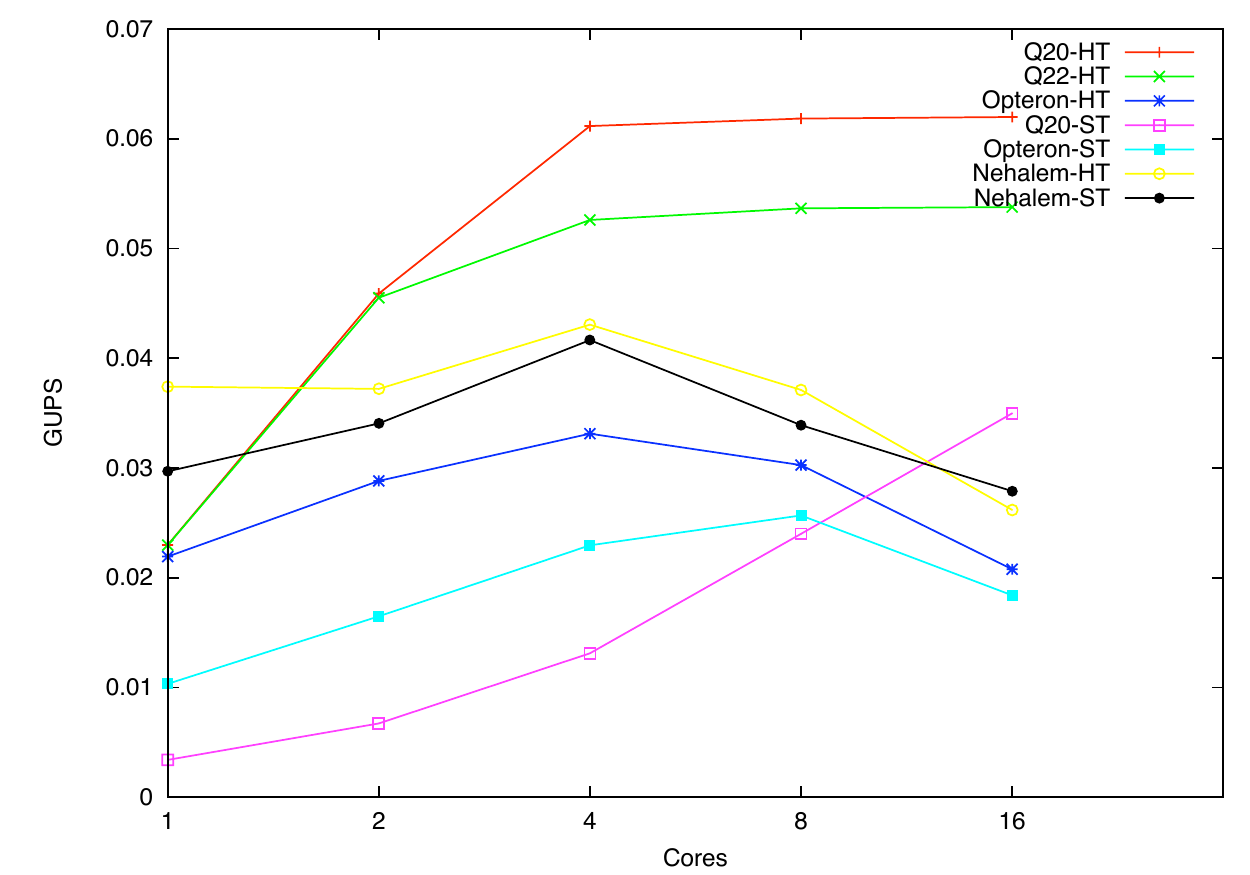,width=3.5in}
\caption{GUPS of different Platforms,varying core/thread number (HT: huge,TLB, ST:Small tlb)  }
\end{figure}

The QS20 has about 15\% higher GUPS than QS22 version. This shows XDR memory is good at interleaving. The opteron platform gets a max 0.033 GUPS, about a half of QS20. It should be noticed that on Opteron 16-core has a worse result than 4-8 cores. This may due to the limited cross-processor bandwidth of Opteron. On the contrary, even the 4K-page QS20 can get highest GUPS with 16 SPUs. The Nehalem platform is similar to opteron, with a max 0.043 GUPS on 4-thread,about 70\% of QS20. 

We can see the effect of TLB page size on Opteron and Nehalem are not as large as Cell/BE for they have more powerful TLB mechanisims. 
 
Overall we can see at least for GUPS, Cell/BE is a better platform than conventional multi-core platforms.

 \section{The implementation of SSCA\#2 over Cell/BE}

The pseudo code of SSCA2 kernel 4 V2.2 is as follows \cite{SSCA_2_Spec,bader_design_2005,bader_parallel_2006}:

\textbf{Input}:	$G(V,E)$ /*$|V|=2^{scale},|E|=8*2^{scale}$*/

\textbf{Output}:	Array $BC[1 \dots n]$

\begin{enumerate}
\renewcommand{\labelenumi}{\arabic{enumi}}

\item \textbf{\textbf{for}} all $v \in V$ in parallel \textbf{do}
\item \hspace{0.5cm}	$BC[v] \gets 0$;
\item let $V_S \subseteq V$ and $|V_S|=2^{k4approx}$  /*exact vs. approximate*/
\item \textbf{for} all $s \in V_S$ \textbf{in parallel do}
\item \hspace{0.5cm}	$S \gets$ empty stacks;
\item \hspace{0.5cm}	$P[w] \gets$ empty list, $w \in V$
\item \hspace{0.5cm}	$\sigma[t] \gets 0, t \in V; \sigma[s] \gets 1$;
\item \hspace{0.5cm}	$d[t] \gets -1, t \in V; d[s] \gets 0$;
\item \hspace{0.5cm} 	queue $Q \gets s$;
\item \hspace{0.5cm}	\textbf{while} $Q \ne  \Phi$ \textbf{do} 
\item \hspace{1cm}	dequeue $v \gets Q$;
\item \hspace{1cm}	push $v \to S$ ;
\item \hspace{1cm}	\textbf{for} each neighbor $w$ of $v$ \textbf{in parallel do}
\item \hspace{1.5cm}		\textbf{if} $d[w]<0$ \textbf{then}
\item \hspace{2cm}			enqueue $w \to Q$;
\item \hspace{2cm}			$d[w] \gets d[v] + 1 $;
\item \hspace{1.5cm}		\textbf{if} $d[w]=d[v]+1$ \textbf{then}
\item \hspace{2cm}			$\sigma[w] \gets \sigma[w] + \sigma[v]$;
\item \hspace{2cm}			append $v \to P[w]$;				
\item \hspace{0.5cm}	$\delta[v] \gets 0, v \in V$;
\item \hspace{0.5cm}	\textbf{while} $S \ne \phi$ \textbf{do}
\item \hspace{1cm}		pop $w \gets S$;
\item \hspace{1cm}		\textbf{for} $v \in P[w] $ \textbf{do}
\item \hspace{1.5cm}			$\delta[v] \gets \delta[v] + 
					\frac {\sigma[v]} {\sigma[w]}(1+\delta[w])$;
\item \hspace{1cm}		\textbf{if} $w \ne s$ \textbf{then}
\item \hspace{1.5cm}			$BC[w] \gets BC[w] + \delta[w]$;					
\end{enumerate}

Loop 10-19 is the BFS expansion process; loop 21-26 is the back trace process.

Our implementation uses nearly the same process flow and data structure of the C/OpenMP version. We start to distribute workload on step 11 and 22. The dynamic stack Q is divided evenly to all SPUs, then each SPU will check their part of Q. It can be seen that the step 16 and 18 are global update operations that need atomic operation to assure the consistency. Using atomic instruction of Cell/BE, the two updates can be done in a single 128-byte \texttt{getllar}-check and update-\texttt{putllc} operation. The porting is straightforward at first. 

To get better performance, 3 techniques were used according to the feature of Cell/BE:

\subsection{Dynamic load balancing}

The workload is distributed on step 11 and 22. We take step 11 as an example. Although workload is divided evenly according to Q, the real workload depends on the total number of neighbors of each vertex in step 13-19 as well as the topology of the graph. These could not be acquired before work partitioning. In fact, the scale-free feature of the graph increases the unbalance of workload: the neighbors of a vertex varying from 0 to thousands. 
So we adopt a dynamic load balancing mechanism: each SPU only allocates a small number of vertexes from Q each time, and only reallocates after finished current work. Since allocation needs synchronization also, allocating one by one is not acceptable. In our experiments, this mechanism enhanced the performance by at least 15\%.

\subsection{Prefetching use clustered DMA and DMA-list.}

The Cell/BE SPU has no local cache and no hardware prefetching mechanism. Clustering data access have to be done by hand. However, the program has many steps with data dependences. For example, step 12-19 can be split to following steps:  
\begin{description}
\item 1) Load $v$ from $Q$
\item 2) Load 1 neighbor $w$ of $v$, load weight of edge $\langle v,w \rangle$
\item 3) Check $w$ and weight$\langle v,w \rangle$ 
\item 4) Load $d[w]$ , load $\sigma[w]$ 
\item 5) Check $d[w]$ , update $\sigma[w]$ 
\item 6) Append $Q$, append $P[v]$
\end{description}

Each step is depending on the data or condition from previous step. If single word DMA operation were used, then most time would be wasted on waiting for last DMA to complete. So pre-fetch and post-write buffers for each data were used. Due to the dynamic size of different data variables, this does increase the programming complexity quite a bit. In 4), a DMA-list must be used since {w} is scattering across the graph. 

The atomic update in 5) prevents batching DMA to be used. We have to do atomic update one-by-one to assure consistency. That remains the main delay in the whole program.

\subsection{Software pipelines}

Even using DMA and DMA-list, there is still much time wasted for waiting memory I/Os. Sometimes a vertex only has 1-2 neighbors that make clustering impossible. So we designed a 3-stage software pipeline for step 13 to 19 to reduce the latency further:
\begin{description}
\item{Stage 1)} Load index (${w}$ and $weight\{w\}$) 
\item{Stage 2)} Load scattered data $(d[w], \sigma[w])$)
\item{Stage 3)} Check $d[w]$, do atomic update and post write 
\end{description}
In one loop or time step, stage 1) start loading neighbor of $v_{n+2}$, stage 2) start loading $\sigma[w]$ of neighbor $v_{n+1}$, while the stage 3) is updating $\sigma[w]$ of neighbor $v_n$. Triple buffers are used for three stages. By using the software pipeline, it is no need to do immediate wait-for-complete for all normal DMA operations. This allows more overlap of various DMA operations that can better utilize the DMA capability of MFC.

The scale-free graph adds complexity here again. Since some vertex may have thousands of neighbors, it has to be spread on multi-stage; for some vertex with zero neighbors, an empty stage is inserted. So finally we have an irregular software pipeline with dependency between stages.

The software pipeline works fluent and add at least another 15\% performance. But profiling shows the stage 3 occupies the most time due to the atomic operations that cause stop and wait.

For step 22-24, we use another similiar software pipeline.

To summary, porting SSCA2 is not an easy task. Not only because the algorithm itself is relative complex, the varying workload and data structure size add difficulties for a better performance. 

\section{Performance evaluation of SSCA\#2}
\subsection{SSCA2 behavior on QS20}

We use SSCA\#2 Kernel 4 with scale 18-22, K4Approx= 8. For Scale 22 more than half of the memory on QS22 was used.  
Figure. 9 shows the different run time when varying cores and scale, all axis are in log scale.
\begin{figure}[h]
\centering
\epsfig{file=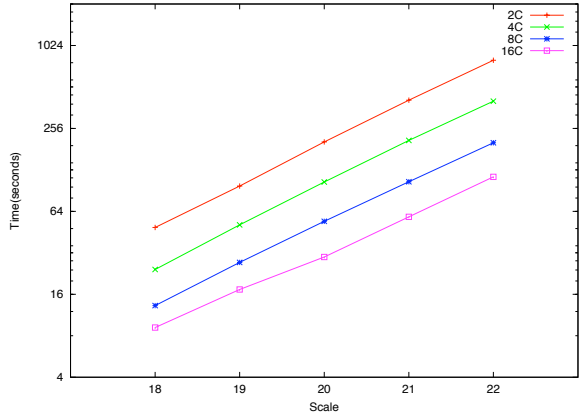,width=3.5in}
\caption{SSCA\#2 on QS20 for varying scale and SPUs }
\end{figure}
For fixed core number we can see a nearly straight line. This means the performance is not changed much for different problem size. 
We can also see a near linear speedup when we add more SPUs, 16-SPU will reach the peak, about 65.8MTEPS.  Compared with above result of GUPS where only 4 SPUs will used up the memory path, we can incur that our implementation has not fully utilized the DMA power of single SPU. The reason may due to the idle delay caused by atomic updates.

\subsection{The internal profiling result}
Using the built-in decrementer of Cell/BE SPU, we analyzed the internal loop of SSCA2 code. Normally the process time ratio for the BFS and the back trace process is about 3.45: 1.

Since a software pipeline was used, all normal DMA operations are asynchronous. It is difficult to tell the exact execution time of each DMA. The exception is atomic update, which a stop-and-wait must be used. The time ratio of the three stages is about 1: 3: 25, while in stage 3, the time period for atomic update occupies about 80\%. In average each atomic update operations elapses about 630ns on QS20, and 550ns on QS22. It should be mentioned that there are still background DMA operations working when the atomic update operation is being executed. So the portion of pure delay brought by atomic operation is undetermined yet.

\subsection{Comparison over different platforms}
We use scale=22, K4approx=8 and varying the core/thread number for different platforms. In this test a Sun T2 5220 (niagara 2) was added. It has 1Ghz  processor, 8 core, 64 physical thread . It has 4 dual-channel FB-DIMM , nearly 60GB memory bandwidth. A special optimized OpenMP version was used.

\begin{figure}[h]
\centering
\epsfig{file=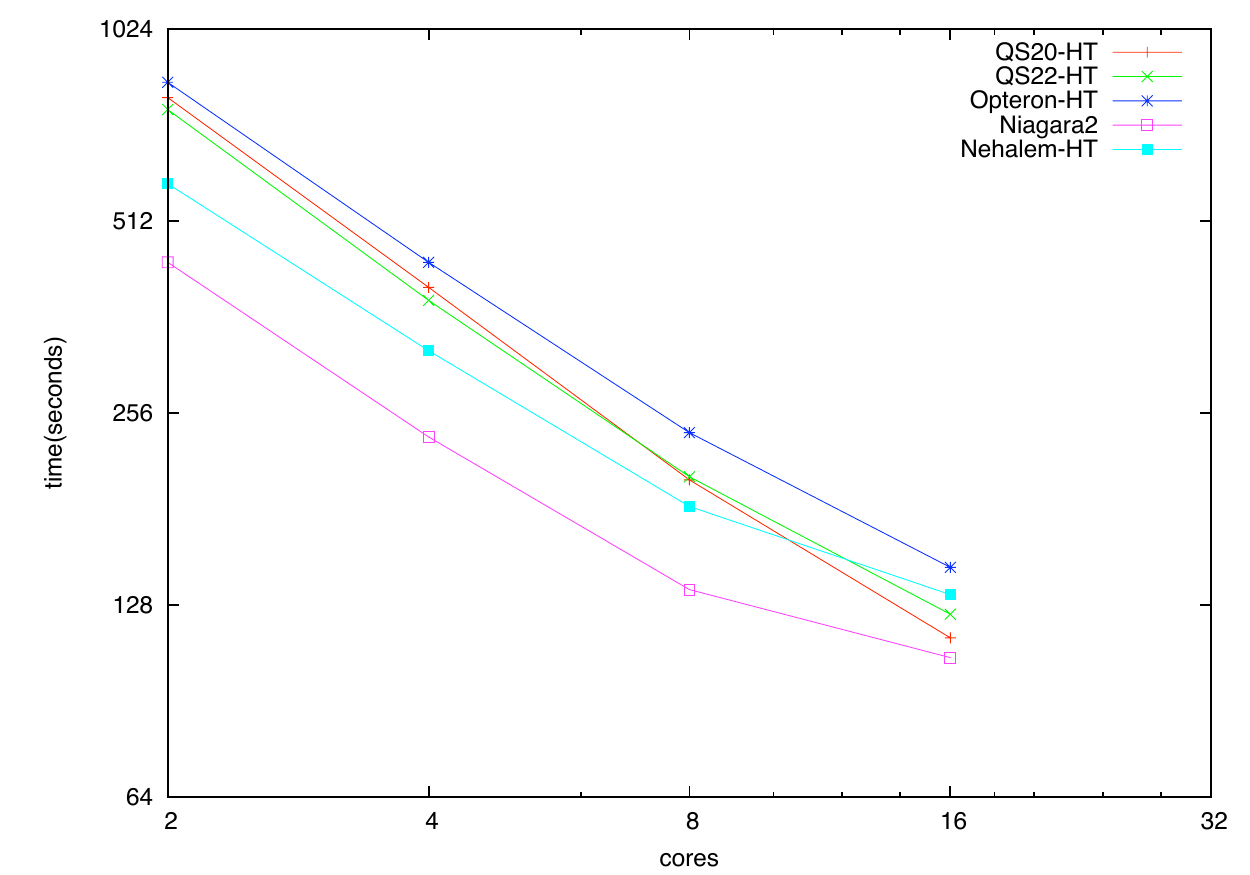,width=3.5in}
\caption{SSCA\#2 K4, Scale=22, varying core/thread number(Niagara should multiply by 4)}
\end{figure}

The best result in these platforms is Sun Niagara 2, about 70.4 MTEPS. The QS20 has a maximum of 65.8MTEPS, about 10\% fast than QS22 , 17\% fast than nehalem and  about 30\% fast than the  Opteron platform, which has the same core/thread number and near memory bandwidth. 

For all platforms we can see the performance keep improving as core/threads increase. It suggests the memory bandwidth are not fully utilized due to more processing logic are needed for such a complex application.

Optimization of SSCA\#2 for highly multithreaded architecture -e.g. SUN Niagara 2- is much more straightforward.  But this work indicated that after applying multiple techniques to hide the memory access latency, the performance of Cell/BE is comparable to Niagara 2 and better than conventional multi-core platforms.

\section{Related works}
Few literatures deal with the memory-intensive application on Cell/BE. Papers \cite{chow_Largefft, bader_fftc:_2007} have studied FFT over Cell, which has a scattered but regular memory access pattern. FFT do have similar feature as GUPS and SSCA2. From these work we got valuable hints include the huge TLB page and fast SPU synchronization.

\cite{kistler_cell_2006} gave a detailed analysis of the communication performance of Cell/BE using micro-benchmark, which encouraged our work on GUPS. They focused on the bus performance and did not give the result when large data set was used. 

\cite{bader_design_2007} presented a software thread idea for list-ranking, which induced us for the GUPS implementation. For the SSCA2, the irregular data size makes much trouble for thread partition. So eventually we used a software pipeline method instead.

\cite{villa_challenges_2007} designed a delicate lock-free BFS algorithm on Cell/ BE. The algorithm depends on a bitmap in SPU's on-chip memory. During the optimization of SSCA2 over Cell/BE, we found the main obstacle was the global atomic update. Each atomic operation will pause the pipeline with idle waiting. However it is not easy to design a lock-free algorithm due to the amount of globally random data updates. The process of SSCA2 need $d[w], \sigma[w]$ and $prev$ set to be updated at the same time during the BFS expansion. These data structure are too large to fit in the on-chip memory. 

In \cite{tan_experienceoptimizing_2008} SSCA2 was porting to an innovative many-core platform, which split cores for memory operations and graph analysis. 

\cite{bader_architectural_2005} discussed how the architectural features of Cray MTA-2 support graph analysis application includes list-ranking and connected components. 

\cite{bader_designing_2006} gave an implementation of BFS over Cray XMT using its unique synchronization feature. 

\cite{BC_no_lock} presented a lock-free algorithm of SSCA2 K4 on multi-core X86 platform based on partitioned data structure. It is still need to check if it is effective on Cell/BE platform also. 

Our implementation of SSCA2 is based on \cite{SSCA_2_Spec,ulrik_brandes_faster_2001,bader_design_2005} and includes the latest change from v2.2.1. We use nearly the same memory data structure and flow  for comparison.

\section{Conclusions}
In this paper, we investigated two memory-intensive benchmarks, GUPS and SSCA2 on the Cell Broad Engine platform. We find both benchmark has good performance on the IBM QS20/22. Compared with 2 conventional multi-core  system with the near memory bandwidth, the GUPS is about 40-80\% fast and SSCA2 about 17-30\% fast. By using dynamic load balancing and software pipeline in SSCA2 we showed that a relatively complex graph analysis application can be port to Cell/BE platform and get a better performance than conventional multi-core platform.

Our works shows that the Cell/BE SPU DMA engine has potential capability for more random accesses, which is restricted by the memory controller; the TLB page size will affect the random access performance greatly on large dataset; the overall memory access performance will be degraded if large amount of atomic operation exists.

There remains an open problem whether there is an efficient lock-free algorithm for SSCA2 to exploit more memory access capability of the Cell/BE platform.

\section{Acknowledgments}
Thanks Kamesh Madduri for providing us the C/OpenMP v2.2.1 version and optimized version for SUN T2 before public announcement. 

%
\bibliographystyle{abbrv}
\bibliography{cell}  

\begin{thebibliography}{10}

\bibitem{HPC_Graph_analysis}
Hpc graph analysis.
\newblock {\em http://www.graphanalysis.org/benchmark/index.html}.

\bibitem{HPCC_challenge}
The hpcc chanllenge benchmark.
\newblock {\em http://icl.cs.utk.edu/hpcc}.

\bibitem{numactl}
numactl and libnuma.
\newblock {\em http://oss.sgi.com/projects/libnuma}.

\bibitem{GUPS_web}
Random access: Gups (giga updates per second).
\newblock {\em from http://icl.cs.utk.edu/projectsfiles/hpcc/RandomAccess/}.

\bibitem{bader_fftc:_2007}
D.~Bader and V.~Agarwal.
\newblock {\em {FFTC:} Fastest Fourier Transform for the {IBM} Cell Broadband
  Engine}, pages 172--184.
\newblock 2007.

\bibitem{bader_design_2007}
D.~Bader, V.~Agarwal, and K.~Madduri.
\newblock On the design and analysis of irregular algorithms on the cell
  processor: A case study of list ranking.
\newblock In {\em Parallel and Distributed Processing Symposium, 2007. {IPDPS}
  2007. {IEEE} International}, pages 1--10, 2007.

\bibitem{bader_design_2005}
D.~Bader and K.~Madduri.
\newblock {\em Design and Implementation of the {HPCS} Graph Analysis Benchmark
  on Symmetric Multiprocessors}, pages 465--476.
\newblock 2005.

\bibitem{bader_parallel_2006}
D.~Bader and K.~Madduri.
\newblock Parallel algorithms for evaluating centrality indices in real-world
  networks.
\newblock In {\em Parallel Processing, 2006. {ICPP} 2006. International
  Conference on}, pages 539--550, 2006.

\bibitem{bader_architectural_2005}
D.~A. Bader, G.~Cong, and J.~Feo.
\newblock On the architectural requirements for efficient execution of graph
  algorithms.
\newblock In {\em Proceedings of the 2005 International Conference on Parallel
  Processing}, pages 547--556. {IEEE} Computer Society, 2005.

\bibitem{bader_designing_2006}
D.~A. Bader and K.~Madduri.
\newblock Designing multithreaded algorithms for {Breadth-First} search and
  st-connectivity on the cray {MTA-2}.
\newblock In {\em Proceedings of the 2006 International Conference on Parallel
  Processing}, pages 523--530. {IEEE} Computer Society, 2006.

\bibitem{ulrik_brandes_faster_2001}
U.~Brandes.
\newblock A faster algorithm for betweenness centrality.
\newblock {\em Journal of Mathematical Socialogy}, 25(2):163--177, 2001.

\bibitem{chow_Largefft}
A.~Chow, G.~C. Fossum, and D.~A. Brokenshire.
\newblock A programming example: Large fft on the cell broadband engine.
\newblock {\em IBM Corp. May 2005}.

\bibitem{IBM_CELLBE}
I.~corp.
\newblock Cell broadband engine architecture, version 1.02.
\newblock {\em http://www.ibm.com/}, 2007.

\bibitem{David_peta}
D.~A.~B. (Ed.).
\newblock {\em Petascale Computing: Algorithms and Applications}.
\newblock Chapman \& Hall/CRC Computational Science Series, 2007.

\bibitem{SSCA_2_Spec}
D.~A.~B. et~al.
\newblock Hpcs scalable synthetic compact applications \#2 graph analysis,
  version 2.2.
\newblock {\em
  http://www.graphanalysis.org/benchmark/HPCS-SSCA2\_Graph-Theory\_v2.2.pdf},
  2007.

\bibitem{kistler_cell_2006}
M.~Kistler, M.~Perrone, and F.~Petrini.
\newblock Cell multiprocessor communication network: Built for speed.
\newblock {\em Micro, {IEEE}}, 26(3):10--23, 2006.

\bibitem{BC_no_lock}
G.~Tan.
\newblock A fine-grained parallel betweenness centrality algorithm without lock
  synchronization.
\newblock In {\em Parallel Processing, 2009. {ICPP} 2009. International
  Conference on}.

\bibitem{tan_experienceoptimizing_2008}
G.~Tan, D.~Fan, J.~Zhang, A.~Russo, and G.~R. Gao.
\newblock Experience on optimizing irregular computation for memory hierarchy
  in manycore architecture.
\newblock In {\em Proceedings of the 13th {ACM} {SIGPLAN} Symposium on
  Principles and practice of parallel programming}, pages 279--280, Salt Lake
  City, {UT,} {USA}, 2008. {ACM}.

\bibitem{villa_challenges_2007}
O.~Villa, D.~Scarpazza, F.~Petrini, and J.~Peinador.
\newblock Challenges in mapping graph exploration algorithms on advanced
  multi-core processors.
\newblock In {\em Parallel and Distributed Processing Symposium, 2007. {IPDPS}
  2007. {IEEE} International}, pages 1--10, 2007.

\end{thebibliography}
%
%
\end{document}